\UseRawInputEncoding 
\documentclass[pre,amssymb,superscriptaddress,showpacs,floatfix,a4paper,twocolumn]{revtex4-1}

\usepackage{graphicx}
\usepackage{bm}
\usepackage{mathptmx}

\usepackage{color}
\usepackage{amsmath}
\usepackage{amssymb}
\usepackage{xcolor,cancel}

\begin{document}

\title{Changing the universality class of the three-dimensional Edwards-Anderson \\ 
spin-glass model by selective bond dilution}

\author{F. Rom\'a}
\affiliation{Departamento de F\'isica,  
Universidad Nacional de San Luis, 
Instituto de F\'isica Aplicada (INFAP), 
Consejo Nacional de Investigaciones Cient\'ificas y T\'ecnicas (CONICET), 
Chacabuco 917, D5700BWS San Luis, Argentina}

\begin{abstract}

The three-dimensional Edwards-Anderson spin-glass model
present strong spatial heterogeneities well characterized by
the so-called {\em backbone}, a magnetic structure that arises as a consequence  
of the properties of the ground state and the low-excitation 
levels of such a frustrated Ising system.  
Using extensive Monte Carlo simulations and finite size scaling, 
we study how these heterogeneities affect the phase transition of the model.
Although, we do not detect any significant difference between 
the critical behavior displayed by the whole system 
and that observed inside and outside the backbone,
surprisingly, a selective bond dilution 
of the complement of this magnetic structure
induces a change of the universality class, 
whereas no change is noted when the backbone is fully diluted.
This finding suggests that the region surrounding the backbone 
plays a more relevant role in determining the physical properties 
of the Edwards-Anderson spin-glass model than previously thought. 
Furthermore, we show that when a selective 
bond dilution changes the universality class of the phase transition, 
the ground state of the model does not undergo any change. 
The opposite case is also valid, i. e., 
a dilution that does not change the critical behavior 
significantly affects the fundamental level.
\end{abstract}

\pacs{75.10.Nr, %Spin-glass and other random models
    75.40.Mg} %Numerical simulation studies

\date{\today}

\maketitle

\section{Introduction}

Both, the singular phenomenology that glassy materials display 
and the enormous technical difficulties 
that must be overcome in order to study them, 
have promoted them as one of the central topics in modern physics.
The most serious attempts to elucidate the physical origin of this behavior 
have led to the development of highly sophisticated experimental, 
theoretical and simulation techniques.    
Although over the years this effort has paid off, 
it is still unclear how, under certain conditions 
relatively simple models are able to exhibit this phenomenon.

Typical glassy models have been studied countless times 
by performing increasingly powerful simulations  
that have characterized with greater precision 
some few aspects of the problem, 
but that have not been enough to get to the bottom of this matter.
At present, however, studying the heterogeneities 
that characterize these complex systems 
seems to be a promising way to understand 
them a little more in depth \cite{Berthier2011}. 
This idea has been exploited in different ways 
trying to infer the basic mechanisms behind the physical behavior observed, 
for example, in colloids \cite{Kegel2000,Weeks2000,Mishra2015,Heckendorf2017},
granular matter \cite{Dauchot2005,Ferguson2007,Avila2014,Hentschel2019}, 
and glasses \cite{Hiwatari1998,Bennemann1999,Berthier2005,Kolton2005,Wang2018,Hoshino2020} 
  
In particular spin glasses, magnetic systems that have both 
quenched disorder and frustration \cite{Binder1986,Fischer1991},
are heterogeneous by nature. 
The paradigmatic Edwards-Anderson (EA) spin-glass model \cite{EA}
present both spatial and dynamical 
heterogeneities that have been extensively studied 
focusing on different aspects of this phenomenon
\cite{Ricci2000,Chamon2002,Castillo2003,Montanari2003,Belletti2009,Alvarez2010,Fernandez2013,Fernandez2016,Martin2017}  
In particular, one of these approaches \cite{Roma2006,Roma2007b,Roma2016} 
relies on the observation that 
this system has a structure called {\em backbone} \cite{Barahona1982}
which, in general, originates as a consequence of the
properties of its ground state and its low-excitation levels \cite{Roma2010b,Roma2013}.  
Extensive simulations have shown that the nonequilibrium 
dynamic behavior displayed within the backbone 
differs qualitatively from what is observed outside of this structure.
Such numerical results suggest that the separation of the system into two components
(the backbone and its complement) is not trivial,
so a suitable {\em backbone picture} could be essential 
to describe the physics of spin glasses.
In addition, it is important to note the importance 
of this structure for other related systems such as 
the $K$-satisfiability model, for which the fraction of backbone spins
is the order parameter that characterizes its critical behavior \cite{Monasson1999}.   

Unfortunately, there are great difficulties to perform such calculations.
On one hand, since the computation of ground state configurations 
of the three-dimensional (3D) EA model 
is a non-deterministic polynomial-time (NP)-hard problem, 
considerable numerical effort is required 
to calculate the backbone of a particular 
realization of the quenched disorder (sample).
As a consequence, only a limited number of samples of small sizes 
can be calculated efficiently.      
On the other hand, although in average approximately $57 \%$ of bonds 
belong to the backbone and the rest to its complement,
for a considerable number of samples these structures can have very different sizes,
i. e., their size distributions are very broad \cite{Roma2010b}.
In addition, both the backbone and its complement percolates (simultaneously),
and are composed by a giant component and several finite clusters.
These factors make it extremely difficult to determine 
which processes dominate physics within these regions.

In this work we focus on studying how these spatial heterogeneities 
affect the critical behavior of the 3D EA model.
Using Monte Carlo simulations we calculate at equilibrium
and for different lattice sizes,  
the correlation length and the spin-glass susceptibility
for the whole system but also for the backbone and its complement.
A finite-size scaling analysis suggests that the critical behavior
is unaffected by such heterogeneities.
However, surprisingly, the universality class of the phase transition
can be changed by a selective bond dilution:
While no changes are observed when the backbone is completely diluted,
in the opposite case in which the complement of this structure is removed
we obtain a different set of critical exponents. 
This finding suggests that the region surrounding the backbone 
plays a more relevant role than previously thought 
and therefore we will call it the {\em glass} region.
Furthermore, we show that when a selective 
bond dilution changes the universality class of the phase transition, 
the ground state of the system does not undergo any change. 
The opposite case is also valid, i. e., 
a dilution that does not change the critical behavior 
significantly affects the fundamental level.
 
The outline of the paper is as follows.  
In Sec.~\ref{ModGS} we present the Edwards-Anderson spin-glass model 
and we show how its spatial heterogeneities can be well
characterized by the backbone and the glass.   
Then, in Sec.~\ref{CriBeh} a numerical study of the critical behavior 
within each of these regions is presented,
both for undiluted and diluted lattices.
The effects of a selective bond dilution
on ground-state energies are analyzed in Sec.~\ref{GSene}.
Finally, Sec.~\ref{SumCon} is devoted to summarizing 
the results and conclusions obtained in this work.

\section{The Edwards-Anderson spin-glass model
and its spatial heterogeneities \label{ModGS}}

In the 3D EA spin-glass model \cite{EA}, 
a set of $N$ Ising spins $\sigma_i = \pm 1$ are placed in 
a cubic lattice of linear dimension $L$ ($N=L^3$). 
Its Hamiltonian is
\begin{equation}
H = - \sum_{(i,j)} J_{ij} \sigma_{i} \sigma_{j},
\label{ham}
\end{equation}
where $(i,j)$ indicates a sum over the six nearest neighbors. 
The coupling constants or bonds, $J_{ij}$'s, are independent random variables 
drawn from a distribution with mean value zero and variance one.
Here, we use a bimodal distribution, i.e., $J_{ij}=\pm 1$ with equal probability.
In order to minimize finite-size effects 
we take periodic boundary conditions in all directions.  

This model has a highly degenerate ground state \cite{Kirkpatrick1977,Hartmann2000}.
For a single sample it is possible to identify the so-called {\em rigid} bonds 
which do not change their state (satisfied or frustrated) 
in all its ground-state configurations \cite{Barahona1982}.
Those bonds form the backbone while the complementary set, 
the {\em flexible} bonds, composes the glass region.
Using the algorithm proposed in Ref.~\cite{Roma2010b},
we have calculated both structures 
for $10^4$ samples for each size $L=4$ and $L=6$,
$10^3$ for $L=8$, and $320$ for $L=10$. 
In addition, to calculate different observables at equilibrium
we use a parallel tempering algorithm \cite{Geyer1991,Hukushima1996}. 
Details of the simulations are given in Appendix \ref{AppA}.

\begin{figure}[t!]
\includegraphics[width=\linewidth,clip=true]{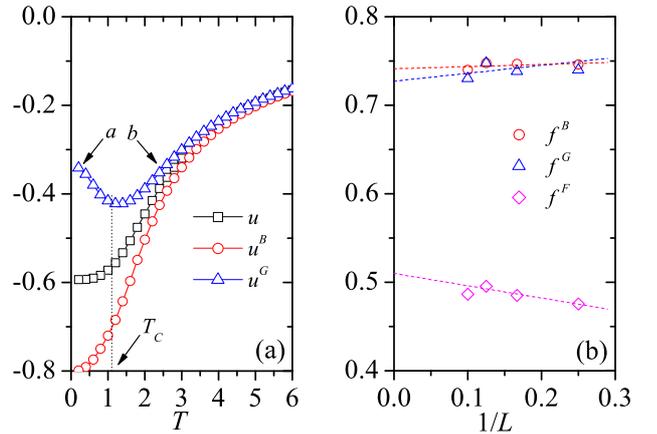}
\caption{(Color online) (a) Average energies per bond $u$, $u^B$, and $u^G$,
as function of $T$.  Arrows indicate the critical temperature
and two particular points on the $u^G$ curve, $a$ and $b$, 
located below and above $T_c$.
(b) Average fractions of spins $f^B$, $f^G$, and $f^F$, 
as function of $1/L$.
}
\label{figure1}
\end{figure}

At low temperatures spatial heterogeneities affect almost any observable. 
For example, Fig.~\ref{figure1} (a) shows 
the average energies per bond $u$, $u^B$, and $u^G$,
as function of temperature $T$ for, respectively, the whole system, 
the backbone, and the glass, for samples of $L=10$. 
Note that the curve of $u^G$ display a minimum at approximately
the critical temperature $T_c=1.1019(29)$ \cite{Baity-Jesi2013}
and $u^G > u > u^B$ for finite $T$, 
which evidence that the glass region concentrates 
most of the frustration of the system. 
We indicate with arrows two particular points, $a$ and $b$, 
to show that it is possible to have the same value of $u^G$ at
temperatures, respectively, below and above $T_c$,
one of the reasons it was assumed (wrongly) that this region 
is in a paramagnetic phase for $T>0$ \cite{Roma2010b}.

Unlike bonds, separating the spins into groups may not be a trivial task.
Typically two sets are chosen, the solidary spins which maintain their 
relative orientation in all configurations of the ground state
and are connected by rigid bonds to each other,  
and the remaining ones that are called non-solidary spins. 
Although the backbone and the glass regions 
have roughly the same number of bonds, 
this separation produces two sets with very different fractions of spins:
$77 \%$ and $23 \%$ of, respectively, solidary and non solidary spins. 
Such a rule was chosen because, {\em a priori}, the backbone 
was considered the most important structure in the system.

Here, assuming both the backbone and glass are of equal importance,
we use these structures to separate the spins into two groups.     
We call $\Omega^B$ ($\Omega^G$) the set of spins connected 
to almost a rigid (flexible) bond, 
where the superscript $B$ ($G$) indicates that such region is dominated 
by the backbone (glass).
Since some spins are connected to both rigid and flexible bonds
these two sets intersect, $\Omega^F=\Omega^B \bigcap \Omega^G$,
where now the superscript $F$ denotes the frontier between
these structures.  
 
In Fig.~\ref{figure1} (b) we can see the average fractions of spins that belong 
to the sets $\Omega^B$ ($f^B$), $\Omega^G$ ($f^G$), and $\Omega^F$ ($f^F$), 
as a function of $1/L$. 
A naive extrapolation to the thermodynamic limit indicates that  
these quantities tend approximately to   
$f^B \sim 0.74$, $f^G \sim 0.73$, and $f^F \sim 0.51$,
which shows that the mean numbers of spins in the backbone and glass regions
are very similar, and the frontier has half of the spins of the system, 
i. e., both structures interpenetrate each other 
sharing a region whose size is proportional to $N$.

\section{Critical behavior \label{CriBeh}}

After having separated the spins in different sets as described above, 
it is possible to analyze other observables within each of these regions.
As usual, to study the critical behavior of the model
we have calculated the Binder cumulant of the overlap order parameter
and the correlation length (which also depends on the overlap between 
two replicas of the system, see below).
The advantage of using these observables is that they allow one to determine 
the existence of a continuous phase transition regardless 
of the type of symmetry that is broken at the critical point.
Nevertheless, we will only show results of the correlation length because, 
as is well known, unlike the Binder cumulant for small lattice sizes 
it allows one to determine the critical point with greater precision \cite{Katzgraber2006}.
 
The correlation length $\xi^x$ is defined as \cite{Palassini1999}
(in cases where a given quantity is not calculated over the whole system, 
we use a superscript $x$ to indicate the region over which it is evaluated)
\begin{equation}
\xi^x=\frac{1}{2 \sin(|\mathbf{k}_{\textrm{min}}|/2)} 
\Bigg[ \frac{\chi^x(0)}{\chi^x(\mathbf{k}_{\textrm{min}})} - 1 \Bigg]^{1/2},
\label{xi}
\end{equation}
where $\mathbf{k}_{\textrm{min}}=(2\pi/L,0,0)$ is the smaller nonzero wave vector
and $\chi^x(\mathbf{k})$ is the wave vector dependent spin-glass susceptibility,
\begin{equation}
\chi^x(\mathbf{k})=\frac{1}{N_{\Omega^x}} \sum_{i,j \in \Omega^x} 
[\langle q_i q_j \rangle_T]_{\textrm{av}} \ 
e^{\textrm{i} \mathbf{k} \cdot (\mathbf{r}_i-\mathbf{r}_j)  } .
\label{Chi_k}
\end{equation}
Here, $q_i=\sigma_i^\alpha \sigma_i^\beta$ is the single spin overlap 
between two replicas of the system $\alpha$ and $\beta$,
$N_{\Omega^x}$ is the number of spins of region $\Omega^x$,
and $\mathbf{r}_i$ is the vector of the position of the $i$-th spin. 
$\langle \cdots \rangle_T$ and $[ \cdots ]_{\textrm{av}}$ represent, respectively,
the thermal and disorder averages.  
The correlation length divided by $L$ is a dimensionless 
quantity which scales as \cite{Katzgraber2006}
\begin{equation}
\frac{\xi^x}{L}=\tilde{S^x}[L^{1/\nu^x} (T-T_c^x)/T_c^x],
\label{FSSxi}
\end{equation}
where $\tilde{S^x}$ is a universal scaling function, 
$T_c^x$ is the critical temperature, and $\nu^x$ is a critical exponent \cite{nota1}.  
If the system experiences a phase transition, according to Eq.~(\ref{FSSxi}) 
the curves of $\xi^x/L$ for different lattice sizes should intersect at $T_c^x$.    
 
Given the limitations under which we carry out this study, i. e.,
we only dispose of a limited number of samples of small sizes,
it is not appropriate to analyze the critical behavior of the model
by a standard scaling technique.
In doing this, one obtains critical parameters 
(the critical temperature and critical exponents)
whose values do not rival those already estimated previously
using systems of much larger sizes \cite{Baity-Jesi2013}. 
Instead, we will focus on checking 
(trying to get good data collapses of the curves)
whether the phase transitions 
observed under different conditions are compatible  
with the universality class of the 3D EA model.
If, for a given case, it is not possible to do this well, 
then we will calculate a suitable set of critical parameters 
for that particular situation.

\begin{figure}[t!]
\includegraphics[width=6.7cm,clip=true]{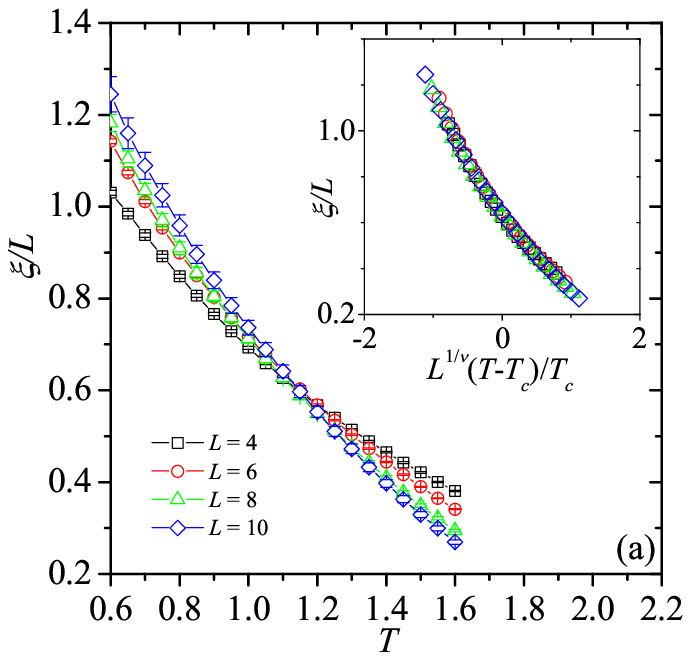}
\includegraphics[width=6.7cm,clip=true]{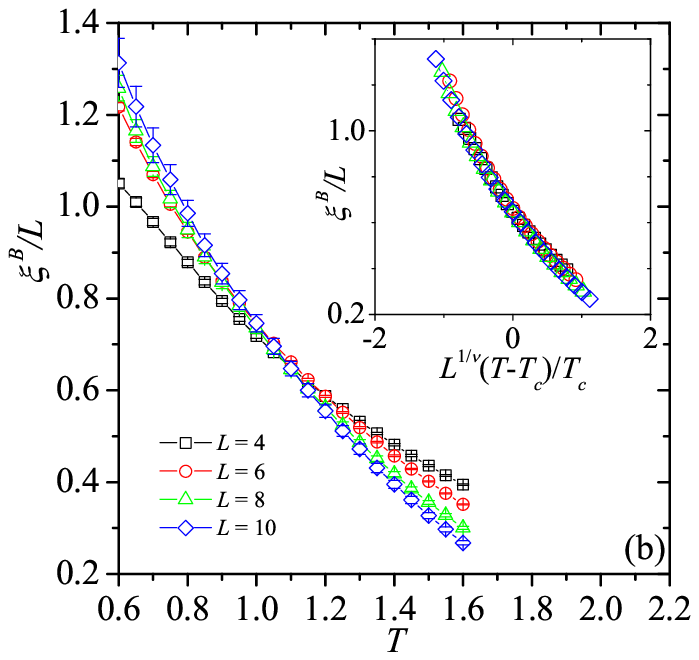}
\includegraphics[width=6.7cm,clip=true]{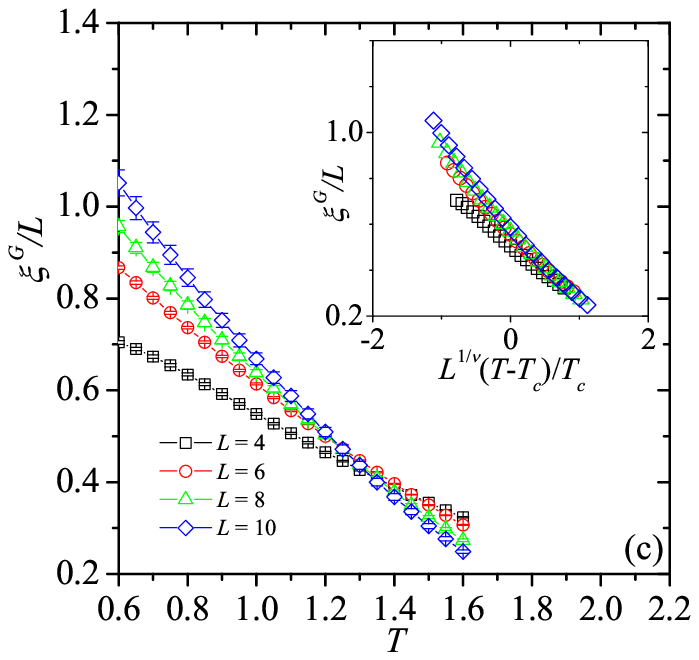}
\caption{(Color online) Correlation length function divided by $L$
as function of $T$ for (a) the whole system, and for  
(b) the backbone and (c) glass regions.
Insets show the corresponding data collapses performed
according to Eq.~(\ref{FSSxi}) using the critical parameters
$T_c=1.1019(29)$ and $\nu=2.562(42)$ \cite{Baity-Jesi2013}.}
\label{figure2}
\end{figure}

Figure \ref{figure2}(a) shows the correlation length calculated 
for the whole system, $\xi$, for different lattice sizes as indicated.
The curves intersect at approximately the true critical temperature and,
using precise values of $T_c=1.1019(29)$ and $\nu=2.562(42)$ 
taken from recent literature \cite{Baity-Jesi2013},
we obtain a good data collapse (see inset).   
This example shows that, despite the limitations of our calculations 
(performed for few samples of very small sizes), 
it is still possible to study the critical behavior of the model
with a certain degree of accuracy.   

Now, we focus on the backbone and glass regions. 
Figure \ref{figure2}(b) shows that the curves 
of $\xi^B$ are very similar to those calculated for the whole system,
and a good data collapse can be obtained using 
the same critical parameters as before (see inset).
For the glass region, however, we do not obtain a result 
as robust as the previous one, see Fig.~\ref{figure2}(c).
Each pair of curves of $\xi^G$ calculated for two consecutive sizes
intersect at a temperature slightly higher than $T_c$,
and their crossing point slowly moves towards lower 
temperatures as the system size increases.
The scaling plot shown in the inset, performed again 
using $T_c=1.1019(29)$ and $\nu=2.562(42)$,
does not allow one to achieve a good data collapse. 
This suggests that the results obtained for the glass region 
are probably affected by very strong finite-size effects.   
 
\begin{figure}[t!]
\includegraphics[width=6.7cm,clip=true]{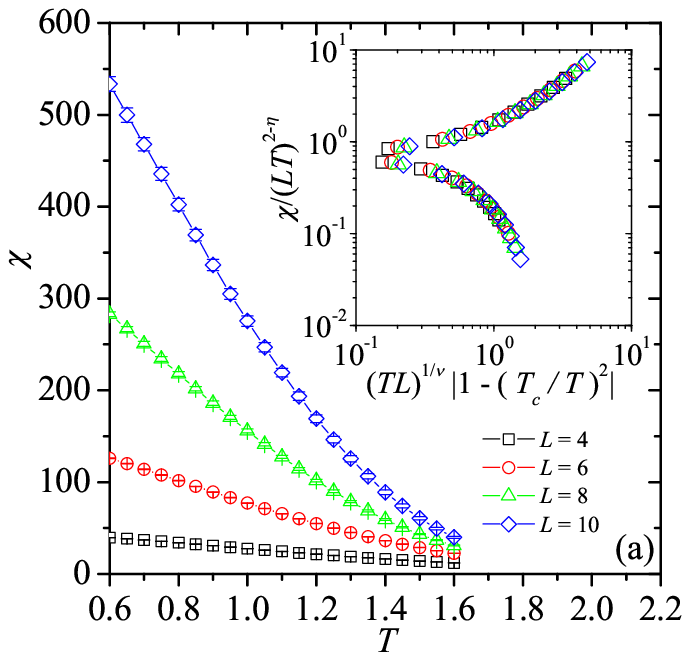}
\includegraphics[width=6.7cm,clip=true]{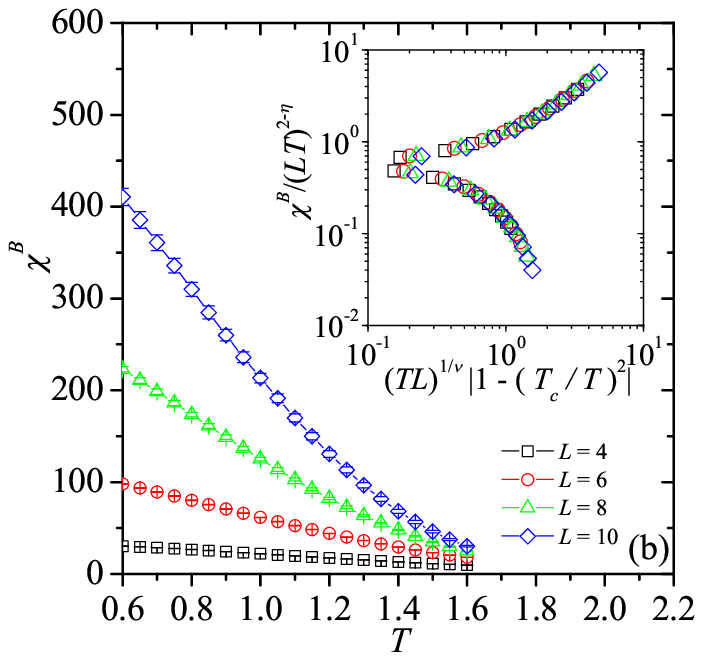}
\includegraphics[width=6.7cm,clip=true]{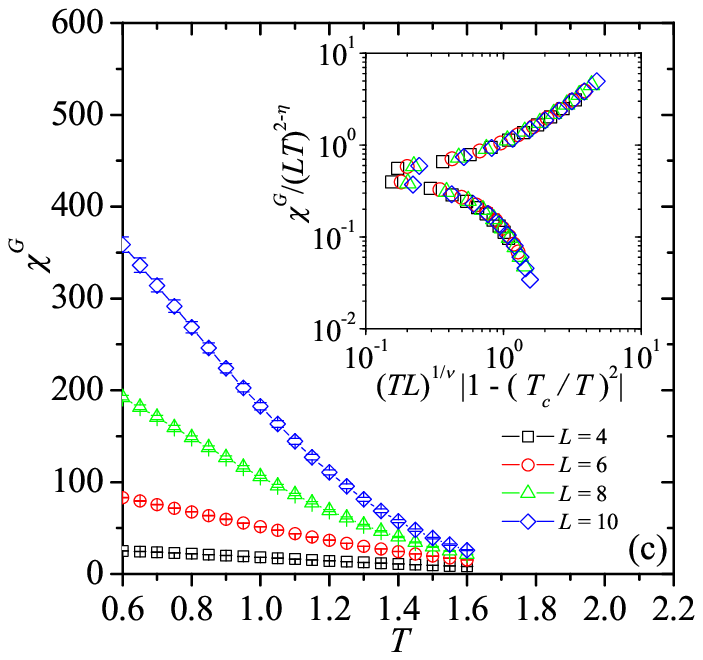}
\caption{(Color online)  Spin-glass susceptibility
as function of $T$ for (a) the whole system, and for  
(b) the backbone and (c) glass regions.
Insets show the corresponding data collapses (see text).}
\label{figure3}
\end{figure}

In order to determine with certainty 
the universality class of a phase transition, 
it is necessary to analyze the scaling of 
a second observable that depends on an independent critical exponent.
Therefore, we calculate the spin-glass susceptibility, 
$\chi^x\equiv \chi^x[\mathbf{k}=(0,0,0)]$,
for each of the regions considered as shown in Figs.~\ref{figure3}(a), 3(b), and 3(c).  
Although for the whole system and the backbone we have obtained 
good data collapses of the correlation length 
using a conventional scaling Eq.~(\ref{FSSxi}),
for the susceptibility we choose an extended scaling scheme \cite{Campbell2006}  
\begin{equation}
\chi^x=(LT)^{2-\eta^x} \tilde{C^x}[(LT)^{1/\nu^x} (1-(T_c^x/T^x)^2)],
\label{ExtScal}
\end{equation}  
which is more appropriate for dealing with samples of small sizes and, 
in addition to $\nu^x$, depends on the critical exponent $\eta^x$.
Using the critical parameters of the 3D EA model, 
$T_c=1.1019(29)$, $\nu=2.562(42)$, and $\eta=-0.3900(36)$ \cite{Baity-Jesi2013},
we obtain excellent data collapses for all regions, 
and in particular for the glass one, see insets in Figs.~\ref{figure3}(a), 3(b), and 3(c).
Thus, we conclude that, despite what is observed in the Fig.~\ref{figure2}(c),
probably the critical behavior is the same 
in each of the regions in which we have divided the system. 

The previous results seem to suggest that the spatial heterogeneities 
we are considering are not closely related to 
the critical behavior of this system.
This conclusion, however, is not entirely correct.
As we shall see below, a selective bond dilution 
allows us to unveil surprising features of  
the backbone and glass regions,
otherwise impossible to detect in a simulation 
that does not take into account such structures.

\begin{figure}[t!]
\includegraphics[width=6.7cm,clip=true]{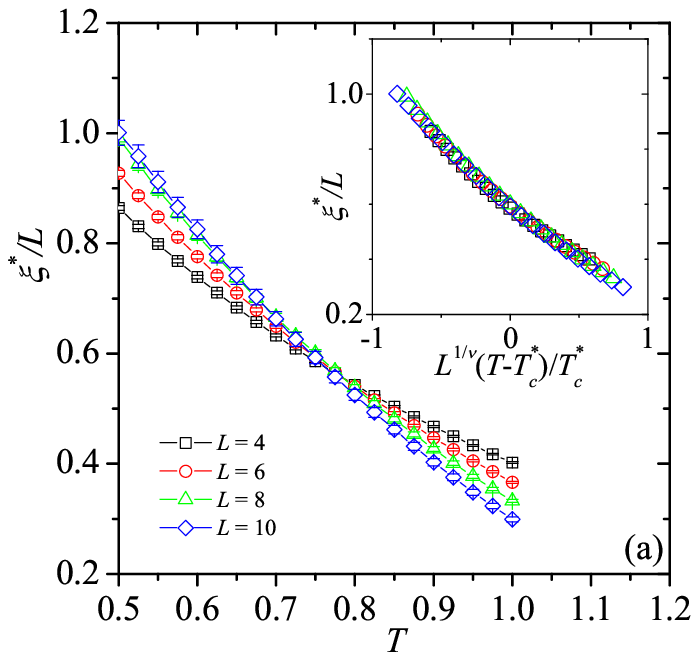}
\includegraphics[width=6.7cm,clip=true]{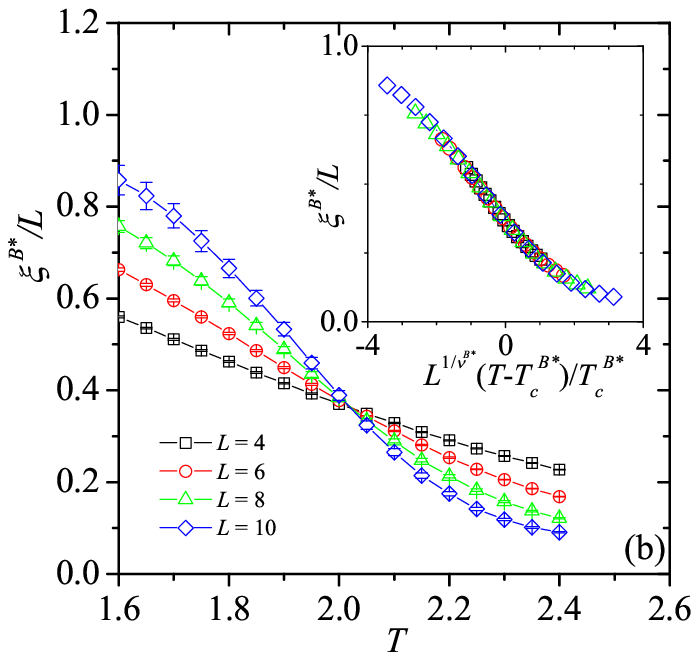}
\includegraphics[width=6.7cm,clip=true]{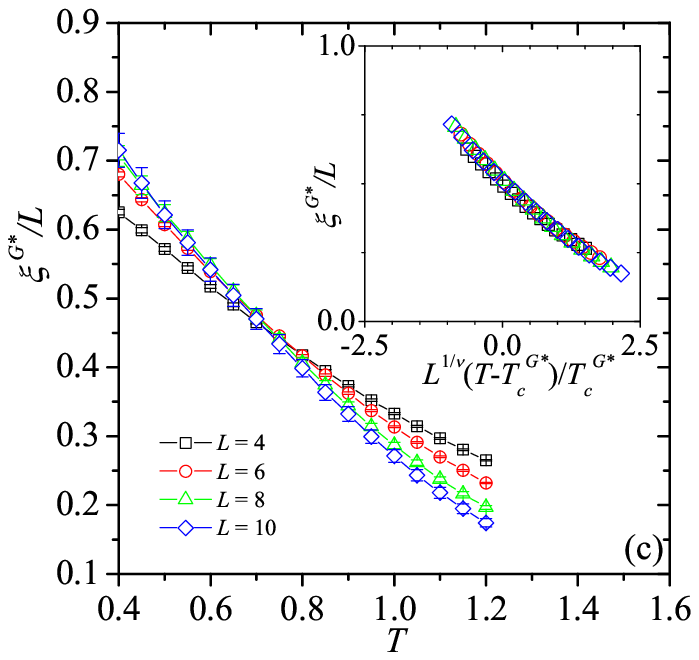}
\caption{(Color online) Correlation length function divided by $L$
as function of $T$ for a random bond dilution (a) of $50 \%$,
and for a selective bond dilution of (b) the glass and (c) backbone regions.   
Insets show the corresponding data collapses (see text).}
\label{figure4}
\end{figure}

\begin{figure}[t!]
\includegraphics[width=6.7cm,clip=true]{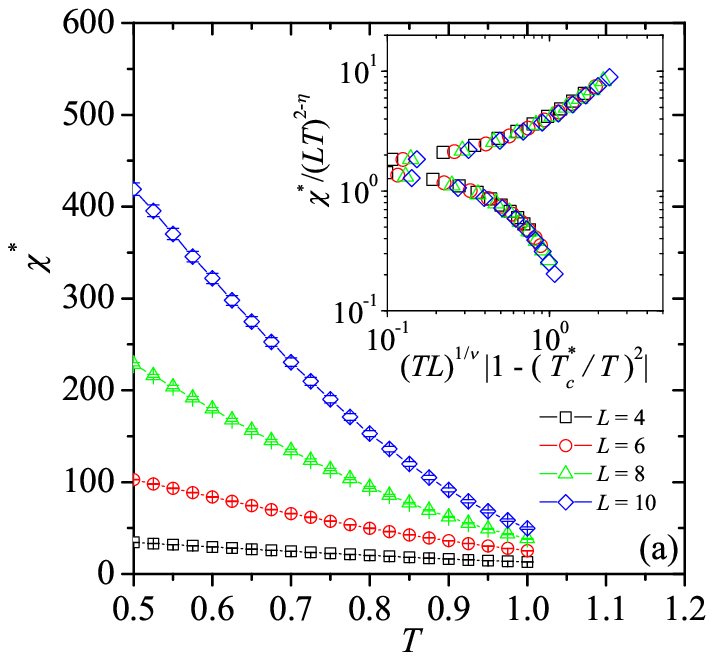}
\includegraphics[width=6.7cm,clip=true]{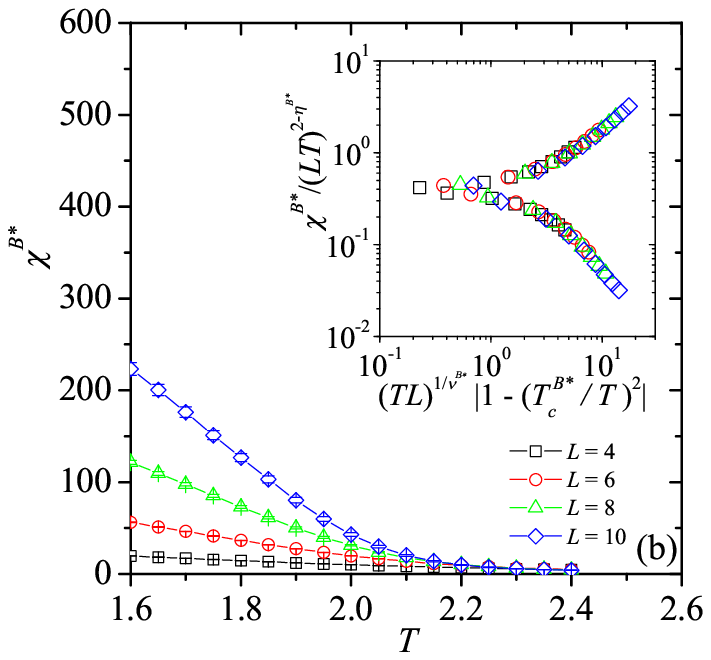}
\includegraphics[width=6.7cm,clip=true]{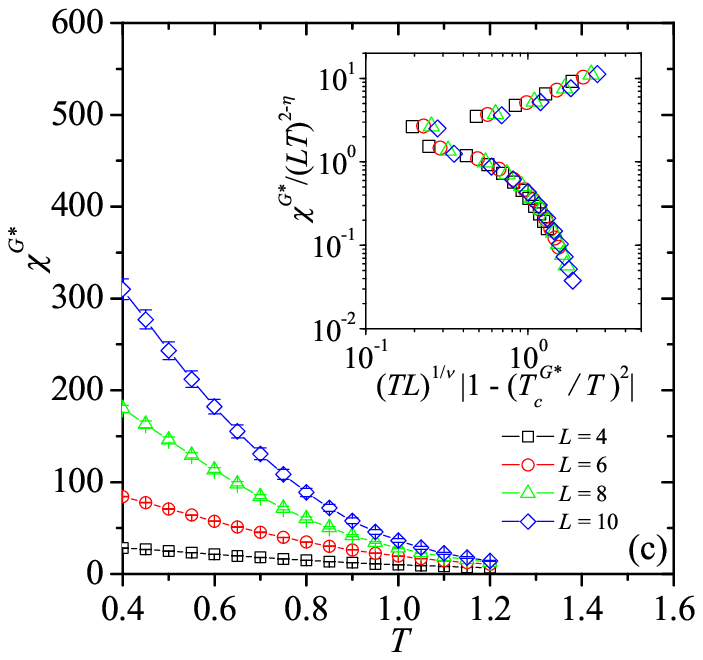}
\caption{(Color online) Spin-glass susceptibility
as function of $T$ for a random bond dilution (a) of $50 \%$,
and for a selective bond dilution of (b) the glass and (c) backbone regions.   
Insets show the corresponding data collapses (see text).}
\label{figure5}
\end{figure}

First, for comparison purposes,
we calculate the correlation length
and the spin-glass susceptibility 
for the 3D random bond-diluted EA spin-glass model,
$\xi^{\ast}$ and $\chi^{\ast}$,
Figs. ~\ref{figure4}(a) and \ref{figure5}(a), respectively. 
We use the same lattice sizes and number of samples as before,
with $50 \%$ of dilution. 
In Fig.~\ref{figure4}(a) we can observe that the curves of $\xi^{\ast}/L$ 
cross at $T_c^{\ast}=0.75(1)$ and,
using the critical exponents $\nu=2.562(42)$ and $\eta=-0.3900(36)$, 
good data collapses are obtained for this quantity (see inset)
and for the spin-glass susceptibility, see inset in Fig.~\ref{figure5}(a).   
This numerical experiment corroborates something that is well known, 
that a random bond dilution does not change the universality class 
of the 3D EA spin-glass model \cite{Hasenbusch2008}.

Surprisingly, a selective bond dilution 
is capable of inducing a change of universality.
Figure \ref{figure4}(b) shows the correlation length curves
calculated for the backbone region, $\xi^{B\ast}$,
obtained after diluting the entire glass region.
On one hand, we observe a cross at $T_c^{B\ast} = 2.02(1)$,
a critical temperature higher than that of the undiluted system. 
This is an expected result, since by removing the region 
with the greatest frustration of the system, 
the phase should be more stable in terms of energy 
and the critical temperature should rise.
But, on the other hand, even more important is
that when using the critical exponents of the
3D EA spin-glass model, we obtain very bad scaling plots
that strongly suggesting that dilution produced a change in the universality class. 
Performing a careful statistical analysis of the data 
which is described in Appendix \ref{AppB},
we calculate two new exponents for this particular case, 
$\nu^{B\ast}=0.81(1)$ and $\eta^{B\ast}=0.47(1)$.
Using them, good data collapses are obtained 
for the correlation length [see inset in Fig.~\ref{figure4}(b)]
and for the corresponding spin-glass susceptibility, $\chi^{B\ast}$
[see inset in Fig.~\ref{figure5}(b)].
Note that to calculate these values and their error bars 
we use a very limited number of small size samples available, 
so it is not easy to implement a method 
that allows taking into account systematic effects such as scale corrections.
However, by using massive computational resources to improve data statistics
and a standard procedure, 
it is possible that more precise critical exponents can be obtained 
that differ slightly from the values reported here. 

In the opposite case, when we dilute the backbone but keep the glass region, 
we observe in Fig.~\ref{figure4}(c) that 
the correlation length curves, $\xi^{G\ast}$, 
intersect at $T_c^{G\ast}=0.64(2)$.
Unlike the previous case, this critical temperature is lower 
than $T_c$ since the most energetically stable 
region (backbone) has been removed.
Using $\nu=2.562(42)$, the inset shows a much better data collapse  
than that obtained in the undiluted case [inset in Fig.~\ref{figure2}(c)].
To confirm that this phase transition belongs 
to the universality class of the 3D EA spin-glass model,
we make a scaling plot of the susceptibility, $\chi^{G\ast}$,
using the exponent $\eta=-0.3900(36)$,
which again leads to a very good data collapse, see inset in Fig.~\ref{figure5}(c).
In this way, it is justified that we have named 
this part of the system the glass region. 

We emphasize the importance of these findings.
We have been able to show that it is possible to change, 
or leave unaltered, the class of universality of 
the phase transition in the 3D EA spin-glass model 
by performing a selective bond dilution.
This does not appear to be a spurious result since,
as shown in Figs.~\ref{figure4} and \ref{figure5},
the data corresponding to these 
particular cases are not significantly influenced by finite-size effects.
Furthermore, unlike what was done in the insets of these figures,
as tested in Appendix \ref{AppC},
if we try to collapse the curves  
for the backbone (glass) obtained after diluting the entire glass (backbone) region
using the corresponding critical temperature, $T_c^{B\ast}$ ($T_c^{G\ast}$), 
but the set of exponents $\nu=2.562$ and $\eta=-0.39$ 
($\nu^{B\ast}=0.81$ and $\eta^{B\ast}=0.47$),
we obtain very bad scaling plots that reinforce 
the idea that the aforementioned change in the universality class 
is in fact real or at least very likely to be. 

\section{Ground-state energies \label{GSene}}

\begin{figure}[t!]
\includegraphics[width=\linewidth,clip=true]{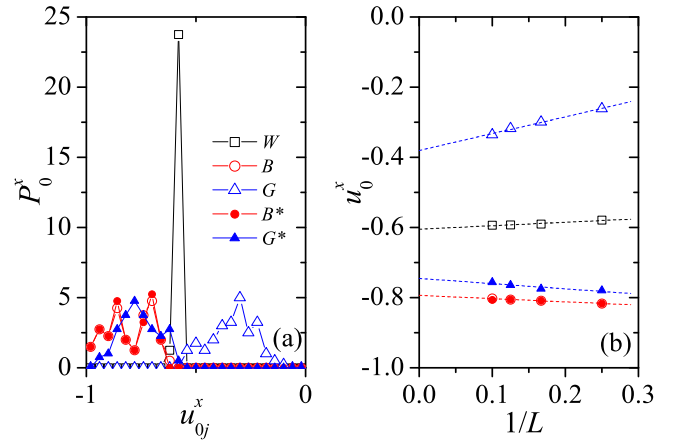}
\caption{(Color online) Probability distribution functions
of ground-state energies per bond for samples of $L=10$,
and for the different regions as indicated.
(b) Disorder average of these energies as function of $1/L$.}
\label{figure6}
\end{figure}

Finally, we observe that a selective bond dilution also affects other properties 
of the system, in particular its fundamental level.  
We calculate the probability distribution function, $P_0^x$,
of ground-state energies per bond, $u_{0j}^x$, 
where as before the superscript $x$ indicates the region
over which this energy is evaluated
(unlike the previous cases, here we use the superscript $W$ to refer to the whole system)
and the conditions under which the calculations are performed 
(the undiluted and diluted cases). 
Figure \ref{figure6}(a) presents the distributions 
that were obtained for samples of $L=10$, 
while the panel (b) shows the disorder average of each energy as function of $1/L$. 
Here, we can see important differences between the main regions of the system.
In fact, a selective bond dilution of the glass region does not change 
the ground state of the backbone 
(the distributions of $u_{0j}^B$ and $u_{0j}^{B\ast}$ are equals),
while in the opposite case an appreciable effect is observed:
the probability distributions of $u_{0j}^G$ and $u_{0j}^{G\ast}$ are very different,
and the latter overlaps appreciably with the corresponding one for the backbone.
Therefore, a selective bond dilution can change 
the universality class of the phase transition of a given region
leaving its ground state unchanged, and vice versa.  

\section{Summary and conclusions \label{SumCon}}

Summarizing, through an extensive analysis 
of the ground state of the 3D EA spin-glass model
we have separated the system in two components,
the backbone and the glass.
We show that the phase transitions observed within each of these regions
have the same class of universality than that has the whole system,
i. e., in the first instance the spatial heterogeneities 
seem not to affect the critical behavior of the model.
However, diluting the glass we observe that
the ground state of the backbone remains unchanged but, more importantly,
we detect that the universality class of the phase transition changes.
In fact, we obtain two critical exponents,
$\nu^{B\ast}=0.81(1)$ and $\eta^{B\ast}=0.47(1)$,
that are very different from those of the 3D EA spin-glass model.
In the opposite case, when the backbone is removed,
we determine that the glass undergoes dramatic 
changes at its fundamental level, 
while the critical behavior remains the same as the undiluted system. 

These results suggest that the critical behavior 
of the 3D EA spin-glass model originates from the interaction 
between two subsystems of very different nature, 
one of which dominates the other
(the glass dominates over the backbone).
Our analysis further reveals that there is no direct but subtle connection between 
the fundamental level of the model and this finite-temperature phenomenon:
Selective dilution has an effect on the ground state 
that is opposite to the one it produces on the critical behavior.
Further studies should be performed to clarify this issue.

Finally, it is important to mention that our results 
could be in conflict with the existence of chaos in temperature \cite{Fernandez2013,Katzgraber2007}.
This phenomenon refers to the fragility of equilibrium configurations 
upon the slightest changes in temperature: 
It is to be expected that the overlap between 
two replicas of the system thermalized at two different but very closed temperatures
tends to zero when the number of spins increases.
In this context, our results showing, for small lattice sizes, 
that there is a connection between the fundamental level 
and the critical behavior (at finite temperature) of the 3D EA model, 
would not make sense at the thermodynamic limit.
We note, however, that 
the backbone of a given sample is obtained from a very particular process:
A bond is rigid (belong to the backbone) if it does not change its state, 
satisfied or frustrated, in all ground-state configurations.
Therefore, this procedure allows one to determine a singular feature of the fundamental level
that is not obtained from a standard equilibrium calculation, 
but arises from a comparison mechanism in which one or 
some particular configurations have the possibility of directly affecting its structure
(in fact, if a bond changes its state in at least one configuration
of the ground state, then said bond will be flexible).
In addition, this information is used only for the purpose 
of separating each sample into two components.
We then conjecture that our results may not be affected 
by the chaotic behavior of this spin-glass model \cite{Roma2013}.
Nevertheless, further investigations similar to those carried out to study 
temperature chaos should be carried out to determine if this is the case.

\begin{acknowledgments}
I acknowledge financial support from 
CONICET (Argentina) under Project No. PIP 112-201301-00049-CO 
and Universidad Nacional de San Luis (Argentina) under Project No. PROICO P-31216. 

\end{acknowledgments}

\appendix
\section{Monte Carlo simulations \label{AppA}}

Monte Carlo simulations are performed using a
parallel tempering algorithm \cite{Geyer1991,Hukushima1996}. 
We use this technique to calculate both 
ground-state configurations \cite{Moreno2003,Roma2009} and 
average values of different observables at equilibrium,
for $10^4$ samples for each size $L=4$ and $L=6$,
$10^3$ for $L=8$, and $320$ for $L=10$. 

This algorithm is implemented as follows.  
We simulate an ensemble of $M$ noninteracting replicas of a system of $N$ spins, 
each one associated to a different temperature in the interval $[T_{min},T_{max}]$ where,
for simplicity, the difference between consecutive temperatures is chosen to be constant. 
A parallel tempering algorithm consists of two routines. 
One of them is a standard Monte Carlo procedure, i. e., 
an attempt to update a random selected spin of the ensemble 
(we randomly choose both a replica and a spin of this replica)
with probability given by the Metropolis rule \cite{Metropolis}.  
The second routine consists of an exchange of 
configurations between two replicas at consecutive temperatures 
which is attempted with the probability defined in Ref.~\cite{Hukushima1996}. 
The unit time (or step) of a parallel tempering algorithm consists of 
$N \times M$ elementary steps of the standard Monte Carlo procedure,
followed by a single trial of replica exchange.

For all cases, with the exception of the curves shown in Fig.\ref{figure1} (a), see below,
the total simulation times (number of parallel tempering steps), $t$, 
required to equilibrate the system are chosen as 
$t=2\times 10^5$ for $L=4$,
$t=4\times 10^5$ for $L=6$,
$t=5\times 10^5$ for $L=8$, and
$t=10^6$ for $L=10$.
We also use between $M=17$ and $M=21$ replicas of the system in each case.
To reach equilibrium under a given condition $x$ (the undiluted and diluted cases),
as usual it is necessary to choose that the highest temperature 
is above the critical one, $T_{max}>T_c^x$.
Once equilibrium is reached, 
the average values of different observables 
are calculated over a time interval of the same length $t$. 
In addition, equilibrium is tested by studying how 
the average values of all observables at all temperatures 
(but especially at $T_{min}$) 
change when $t$ is successively increased by factors of $2$,
requiring that at least the last three results obtained coincide within the error bars.

In the particular case of Fig.\ref{figure1} (a),
the curves are calculated only for $100$ samples of $L=10$ and $M=30$ replicas,
which are simulated between $T_{max}=6.0$ and $T_{min}=0.2$.
Because this last temperature is very low,
to equilibrate we use $t=2\times 10^6$.
We clarify that in this particular case 
the equilibrium was tested taking into account 
only the average energies per bond $u$, $u^B$, and $u^G$,
and therefore other observables such as the corresponding correlation lengths
might not have reached their equilibrium values 
for the lowest temperatures close to $T_{min}$. 
Nevertheless, note that the simulations corresponding 
to Figs.~\ref{figure2} to \ref{figure5}
were carried out independently and for temperature ranges 
different from those of Fig.\ref{figure1} (a).

In addition, it is necessary to find the backbone and glass regions 
of each realization of the quenched disorder.
To determine which are the rigid and flexible bonds of a given sample,
we use a very simple strategy \cite{Ramirez2004,Roma2010b}:
\begin{enumerate}
\item{A ground-state configuration $C$ is calculated and its energy $U_0$ is stored
(to do this, we use a parallel tempering algorithm as explained in Ref.~\cite{Roma2009}).}

\item{Then, a bond $J_{ij}$ of the sample is chosen at random.}

\item{The system being in configuration $C$, 
one of the spins joined by the bond $J_{ij}$, 
i.e. either $\sigma_{i}$ or $\sigma_{j}$, is flipped. 
This flip changes the ``condition" of the bond from 
satisfied to frustrated, or vice versa.}

\item{The orientations of the spins $i$ and $j$ are frozen and,
for this ``constrained" system, 
a new ground-state configuration $C'$ of energy $U'_0$ is calculated.}

\item{If $U'_0 > U_0$, it follows that $J_{ij}$ is a rigid bond
(since we verify that there does not exist a ground-state configuration 
of energy $U_0$ in which this bond appears with its changed condition).}

\item{If $U' = U$, then $J_{ij}$ is a flexible bond
(we find a ground-state configuration of energy $U_0$
in which this bond appears with its changed condition;
this configuration could have been found 
by exhaustively exploring the fundamental level
of the unconstrained system).}

\item{The bond $J_{ij}$ is added to the list of ``checked" bonds, 
and the restrictions over the spins $\sigma_{i}$ or $\sigma_{j}$ are lifted.}

\item{If there are still non-checked bonds, 
a new bond $J_{ij}$ is chosen among them 
and the process is repeated from step 3.}
\end{enumerate}
Ground-state configurations were calculated
with the same parallel tempering algorithm as before,
using the parameters given in Refs.~\cite{Roma2009,Roma2010b}.
Note that in this case such a technique is not used to equilibrate the system
but to quickly reach the fundamental level with a high probability.

\section{Statistical analysis of the data \label{AppB}}

After a selective bond dilution of the glass region, 
we observe that the backbone undergoes a phase transition
that does not belong to the universality class of the 3D EA spin-glass model. 
To determine a suitable set of critical parameters
$T_c^{B\ast}$, $\nu^{B\ast}$, and $\eta^{B\ast}$,
we use a procedure similar to that reported in Refs.~\cite{Katzgraber2006,Matoz2016}.
First, we fit each curve of correlation length  
and susceptibility to a fifth-order polynomial.
Working exclusively with these continuous functions,
we calculate $T_c^{B\ast}$ and $\nu^{B\ast}$ 
looking for the values of these parameters 
that allow us to achieve the best data collapse
of the correlation length using a conventional scaling, Eq.~(\ref{FSSxi}). 
In addition, to determine $\eta^{B\ast}$,
we fix $T_c^{B\ast}$ and $\nu^{B\ast}$ to the values obtained 
previously and we follow a similar procedure for 
the spin-glass susceptibility, i. e.,
we look for the best data collapse of these
curves but now using an extended scaling, Eq.~(\ref{ExtScal}).    
Finally, we calculate the error bars of these new critical parameters
through a bootstrap method by generating $100$ random realizations with replacement
as described in Ref.~\cite{Katzgraber2006}.

\section{Exchanging the critical exponents \label{AppC}}

\begin{figure}[t!]
\includegraphics[width=6.7cm,clip=true]{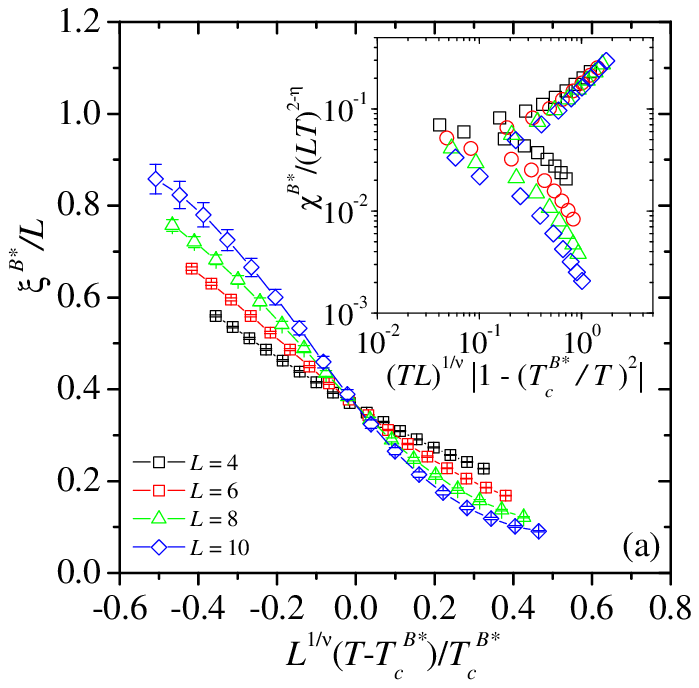}
\includegraphics[width=6.7cm,clip=true]{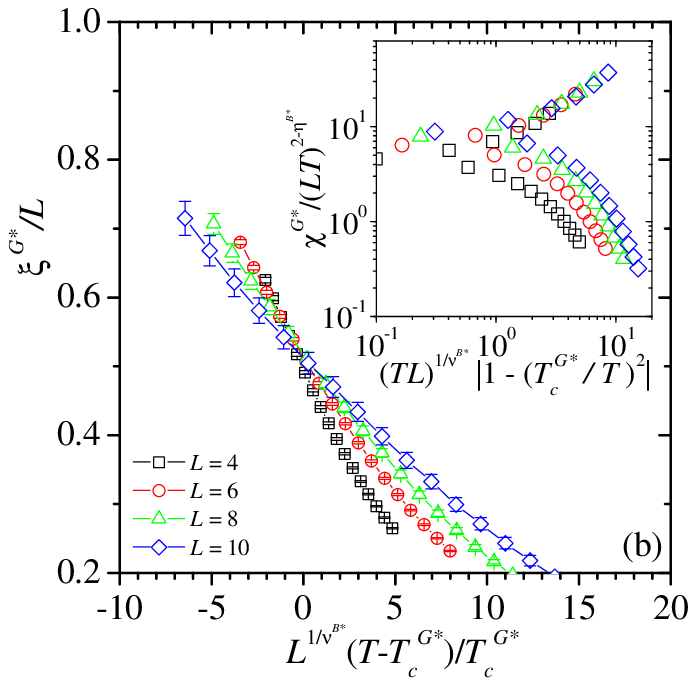}
\caption{(Color online) Scaling plots of the correlation length curves 
(a) $\xi^{B\ast}$ and (b) $\xi^{G\ast}$ obtained using 
the corresponding critical temperatures,
$T_c^{B\ast} = 2.02$ and $T_c^{G\ast}=0.64$, 
and the exponents $\nu=2.562$ and $\nu^{B\ast}=0.81$, respectively.
Insets show the scaling plots of the susceptibility curves (a) $\chi^{B\ast}$
and (b) $\chi^{G\ast}$, calculated with the previous sets of critical parameters, 
and using $\eta=-0.39$ and $\eta^{B\ast}=0.47$, respectively.}   
\label{figure7}
\end{figure}

Figure \ref{figure7}(a) shows the scaling plot 
of the correlation length curves $\xi^{B\ast}$ 
obtained using the corresponding critical temperature,
$T_c^{B\ast} = 2.02$, and the exponent $\nu=2.562$.
Clearly, we can observe the data is far from collapsing 
into a single universal curve.
The same type of test is performed for the correlation length $\xi^{G\ast}$,
in this case using $T_c^{G\ast}=0.64$ and $\nu^{B\ast}=0.81$.
Again we obtain a very bad scaling plot, see Fig. \ref{figure7}(b).
Similarly, the insets in these figures 
show the failed attempt to obtain good data collapses of susceptibility curves,
using the exponent $\eta=-0.39$ for the backbone
and $\eta^{B\ast}=0.47$ for the glass. 

\newpage
%...............................................................................


\begin{thebibliography}{99}
\bibitem{nota1} Since when analyzing a certain region of the system 
the critical temperature and the critical exponents could change, 
to define them a superscript $x$ has been included. 
However, if any of these quantities has the same value 
as the corresponding critical parameter of the 3D EA model, 
then for simplicity the superscript is dropped.
\end{thebibliography}
\end{document}